\documentclass[conference]{IEEEtran}
\usepackage{cite,graphicx,amsmath,amssymb}
\usepackage{subfigure}
\usepackage{citesort}
\usepackage{fancyhdr}
\usepackage{mdwmath}
\usepackage{mdwtab}
\usepackage{balance}
\usepackage{xcolor}
\usepackage{bm}

\usepackage{algorithm}
\usepackage{algorithmic}
\usepackage{multirow}
\usepackage{flafter}
\usepackage[
top    = 1.70cm,
bottom = 1.05in,
left   = 0.63 in,
right  = 0.63 in]{geometry}

\newtheorem{theorem}{Theorem}

\newtheorem{lemma}{Lemma}

\newtheorem{corollary}{Corollary}

\newtheorem{proposition}{Proposition}
\newtheorem{assumption}{Assumption}

\begin{document}

\title{Modeling and Analysis of MmWave Communications in Cache-enabled HetNets
}

\author{
\IEEEauthorblockN{Wenqiang~Yi, Yuanwei~Liu and Arumugam Nallanathan }
\IEEEauthorblockA{Queen Mary University of London, London, UK }
}
\maketitle

\begin{abstract}
  In this paper, we consider a novel cache-enabled heterogeneous network (HetNet), where macro base stations (BSs) with traditional sub-6 GHz are overlaid by dense millimeter wave~(mmWave) pico BSs. These two-tier BSs, which are modeled as two independent homogeneous Poisson Point Processes, cache multimedia contents following the popularity rank. High-capacity backhauls are utilized between macro BSs and the core server. A maximum received power strategy is introduced for deducing novel algorithms of the success probability and area spectral efficiency~(ASE). Moreover, Monte Carlo simulations are presented to verify the analytical conclusions and numerical results demonstrate that: 1) the proposed HetNet is an interference-limited system and it outperforms the traditional HetNets; 2) there exists an optimal pre-decided rate threshold that contributes to the maximum ASE; and 3) 73 GHz is the best mmWave carrier frequency regarding ASE due to the large antenna scale.
\end{abstract}
\vspace{-0.1cm}
\section{Introduction}
\vspace{-0.1cm}
  To support the explosive data traffic of future fifth-generation~(5G) cellular networks, numerous researches~\cite{liu2017caching,7954630,7982794,Zhiguo2015Mag} have paid attention to an innovative framework that densifies the traditional networks with massive small base stations (BSs) or exploiting power domain. However, the improvement of these heterogeneous networks~(HetNets) is mainly restricted to the capacity of the backhauls~\cite{liu2017caching}. A recent study~\cite{6736753} has relaxed the limitation by equipping caches at all BSs to store the most popular files as only the minority of multimedia contents is frequently requested by the majority of customers in the real world. Accordingly, the aforementioned cache-enabled HetNets have been studied in various papers. The primary work~\cite{7445129} analyzed the energy efficiency and throughput of cache-enabled cellular networks with a regular hexagonal grid. Since stochastic geometry is a useful tool to acquire the networks' randomness, modeling a tier of BSs in small cell networks or HetNets with a Poisson Point Process~(PPP) is more accurate than the traditional hexagonal scenario~\cite{heath2013modeling}. Under this condition, the throughput of multi-tier cache-enabled HetNets was discussed in~\cite{liu2017caching}, where macro BSs connected to the core networks via backhauls and small cell devices cached the content through wireless broadcasting. Unfortunately, the further analysis on the impact of backhaul capacity was omitted, which is the key parameter when comparing with the conventional HetNets.

  In addition to the network densification, another key capacity-increasing technology is exploiting new spectrum bands, such as millimeter wave (mmWave)~\cite{8114722}. Two distinctive characteristics of mmWave are small wavelength and the sensitivity to blockages~\cite{rappaport2013broadband,alejos2008measurement}. Thanks to short wavelength, steerable antennas can be deployed at devices to enhance the directional array gain~\cite{rappaport2013broadband}. On the other side, the sensitivity gives rise to severe penetration loss for mmWave signals when passing through building exteriors~\cite{alejos2008measurement}, so it is unrealistic to expect the outdoor-to-indoor coverage from macro mmWave BSs. An ingenious hybrid network is created to solve this issue, where mmWave transmitters contribute to the ultra-fast data rate in short-range small cells, and sub-6~GHz BSs provide the universal coverage~\cite{7493676}. The same with cache-enabled HetNets, stochastic geometry has also been wildly utilized in mmWave networks, where the locations of transceivers were modeled following PPPs~\cite{6932503}. With the aid of such structure, the primary article~\cite{6932503} introduced a stochastic blockage model to represent the actual mmWave communication environment, but the antenna pattern was over-simplified as a flat-top model. Then the authors in~\cite{maamari2016coverage} proposed an actual antenna pattern for increasing the accuracy. Considering the hybrid HetNets, a tractable structure combining mmWave with sub-6 GHz was analyzed in~\cite{7493676}, which performed close to the reality.

 As discussed above, although HetNets with caches have been fully analyzed under traditional sub-6 GHz networks, there is still lack of articles on a hybrid system with mmWave small cells. Since mmWave is able to provide fast data rate in short-distance networks~\cite{8016632}, adopting mmWave into a dense pico tier of cache-enabled HetNets is an promising way to increase the throughput of 5G cellular networks. The other benefit of such hybrid HetNets is no mutual interferences due to applying distinctive carrier frequencies between tiers. These advantages motivate us to create this paper. In contrast to~\cite{liu2017caching}, we introduce fiber-connections between macro BSs and the multimedia server to evaluate the impact of backhaul capacity. Then, due to the employment of mmWave, the propagation environment and antenna beamforming pattern in the small cells are replaced by Nakagami fading and actual antenna arrays, respectively. Moreover, we have compared the performance of various mmWave frequencies in this paper. On the other hand, unlike noise-limited assumption in~\cite{7493676}, we demonstrate that in the dense mmWave networks, the system becomes an interference-limited scenario. With the aid of the content placement, we conclude that the throughput of our system is conditionally decided by the storage capacity. The main contributions are summarized as follows: 1) we discuss the Laplace transform of interference for traditional sub-6 GHz macro cells and mmWave small cells with the actual antenna pattern; 2) the novel algorithms for the success probability and area spectral efficiency (ASE) under \emph{Maximum Received Power} (Max-RP) scheme are derived; 3) our model is an interference-limited system and there is an optimum value of rate requirement for obtaining the maximum ASE; and 4) 73 GHz is the best mmWave carrier frequency under the considered association strategy.
\vspace{-0.1cm}
\section{System Model}\label{System Model}
\vspace{-0.1cm}
\subsection{Network Architecture}
\vspace{-0.1cm}
\begin{figure}[h]
\centering
\includegraphics[width= 3in, height=1.5in]{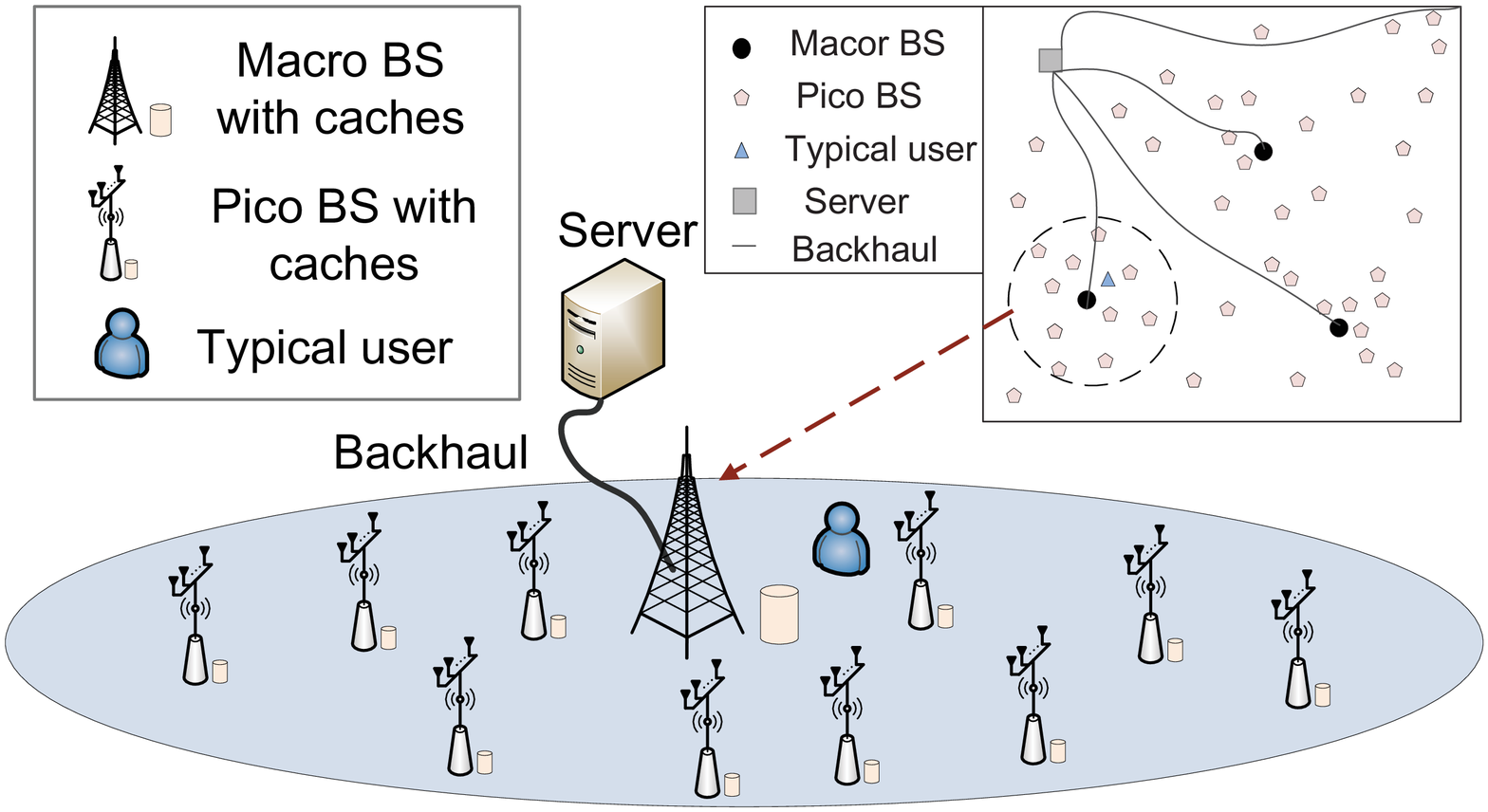}
\caption{Layouts of the proposed cache-enabled two-tier hybrid HetNets with traditional macro cells and mmWave small cells.}
\label{model}
\end{figure}
  In this paper, we present a two-tier cache-enabled hybrid HetNet consisting of numerous macro and pico BSs as shown in Fig.~\ref{model}. The locations of BSs in $i$-th tier are modeled following an independent homogeneous PPP ${\Phi _i}$ with a density $\lambda_i$, where $i=1$ and $2$ represent the macro and pico tier, respectively. In our model, a \emph{typical user} is located at the origin of a plane so that the distribution of distance from the typical user to its nearest BS in $i$-th tier will follow $p_i(r)=2\pi\lambda_i r \exp (-\pi \lambda_i r^2)$. Apparently, the number of pico BSs in the real HetNets is much more than that of macro BSs, so we consider $\lambda_1 \ll \lambda_2$. In order to compare the performance of our proposed networks with the traditional HetNets, we provide a server to supply the less-popular contents. This server connects to each macro BS through a high-capacity wired backhaul.

  Moreover, different carrier frequencies are employed in our system. The macro BSs connect to receivers with traditional sub-6 GHz, while mmWave is utilized between the pico BS and its corresponding user. As a result, there is no interference between these two tiers. However, in the same tier, other active BSs apart from the \emph{corresponding BS} that serves the typical user will be the source of interferences. Additionally, the quantity of users is assumed to be large enough to ensure that all BSs are active when the corresponding BS is transmitting messages to the typical user.
\vspace{-0.1cm}
\subsection{Blockage Model}
\vspace{-0.1cm}
In tier 1, since macro BSs employ sub-6 GHz as its carrier frequency, the path loss law $L_1(\dot{r})$ with a distance $\dot{r}$ is same with that in traditional cellular networks as shown below
\begin{align}
L_1(\dot{r}) = {{C_1}{\dot{r}^{ - {\alpha _1}}},}
\end{align}
where $\alpha_1$ is the path loss exponent and $C_1$ is the intercept for the macro tier.

In tier 2, the communication environment is changed due to the employment of mmWave. More specifically, the transmitting routes of pico BSs are divided into line-of-sight (LOS) group and non-line-of-sight (NLOS) group and each of them has its unique path loss law. Note that the density of the pico tier $\lambda_2$ is huge, which indicates that tier~2 can be visualized as dense mmWave networks. One blockage may obstruct all BSs behind it under this case. Therefore, we consider an LOS ball to model the blockage process. A recent study~\cite{andrews2017modeling} has advocated that such blockage pattern fits the real-world scenarios better than other models. In terms of the NLOS links, the authors in~\cite{6932503} have demonstrated that the impact of NLOS signals is so weak that it can be ignored in mmWave networks, only LOS signals will be considered in this paper.

As a consequence, we define the LOS ball with a radius $R_L$, which represents the departure from nearby obstacles. The probability of LOS links will be one inside the ball and zero outside the area. The path loss law in tier 2 $L_2(\dot{r})$ with a distance $\dot{r}$ is shown as follows
\begin{align}
{L_2}(\dot{r}) = \mathbf{U}({R_L} - \dot{r}){C_2}{\dot{r}^{ - {\alpha _2}}},
\end{align}
where $\alpha_2$ and $C_2$ are the path loss exponent and the intercept of LOS links in tier 2, respectively. $\mathbf{U}(x)$ is the unit step function.
\vspace{-0.5cm}
\subsection{Directional Beamforming}
\vspace{-0.1cm}
In $i$-th tier, we employ antenna arrays composed of $N_i$ elements at all cache-enabled BSs and the transmit power is assumed to be a constant with $P_i$. The uniform linear array formed by $N_2$ antenna elements is utilized at all pico BSs. However, we only consider an omnidirectional antenna pattern at macro BSs ($N_1=1$) and the typical user for tractability of the analysis~\cite{7493676}.

Directional antenna arrays deployed at the pico BSs will supply substantial beamforming gains to compensate the path loss, so the received signal for $i$-th tier at the typical user can be expressed as follows
\begin{align}
y_i =& \sqrt C_i {( {||x_0||} )^{ - \frac{\alpha_i }{2}}}{\textbf{h}^{{x_0}}_i}{\textbf{w}^{{x_0}}_i}\sqrt {{P_i}} {s_{{x_{0}}}} \nonumber\\
&+  \sum\limits_{{x} \in \Phi_i \backslash {x_{0}}} {\sqrt C_i {{( {||x||} )}^{ - \frac{\alpha_i }{2}}}{\textbf{h}^{{x}}_i}{\textbf{w}^{{x}}_i}\sqrt {{P_i}} {s_{{x}}}}  + {n_0},
\end{align}
where interfering BSs are located at $x$ when the typical user is receiving the message from the serving BS at $x_0$. The channel vector from the BS to the typical user and the beamforming vector of the BS in $i$-th tier are denoted by ${\textbf{h}^x_i}$ and ${\textbf{w}^x_i}$, respectively. $n_0$ represents the thermal noise with power $\sigma_i^2$.

 Combining with the aforementioned assumptions, the product of the fading gain and beamforming gain of the BS located at $x$ in $i$-th tier is shown as below~\cite{170204493}
\begin{align}
H^x_i\buildrel \Delta \over =|{\textbf{h}^x_i}{\textbf{w}^x_i}{|^2} = {N_i}|{h_i}{|^2}G_i( {{\varphi _x} - {\theta _x}} ),
\end{align}
where ${h_i}$ is the small fading term for $i$-th tier. ${\varphi _x} $ is the spatial angle of departure (AoD) from the interfering BS to the typical user, and ${\theta _x}$ is the spatial AoD between the BS at location $x$ and its corresponding receiver. $G_i(.)$ is the array gain function. More specifically, an actual array pattern is employed at pico BSs so that $G_2( \omega  ) \buildrel \Delta \over = \frac{{{{\sin }^2}( {\pi {N_i}\omega } )}}{{N_i^2{{\sin }^2}( {\pi \omega } )}}$, where $\omega$ is a uniformly distributed random variable over $[{ - \frac{d}{\lambda }},{ \frac{d}{\lambda }}]$. $d$ and $\lambda$ are the antenna spacing and wavelength, respectively~\cite{170204493}. On the other hand, the array gain function for macro BSs is $G_1( \omega  ) \buildrel \Delta \over = 1$ due to the omnidirectional antenna pattern.
\vspace{-0.5cm}
\subsection{Channel Model}
\vspace{-0.1cm}
Assuming that all active BSs have the full knowledge of the AoD from itself to the typical user, the serving BS will align the antenna beam towards the typical user for achieving maximum directivity gain $G_0=1$. In our model, since the corresponding BS is only interfered by the active BSs located in the same tier, the interference for $i$-th tier can be expressed as follows
\begin{align}
I_i=\sum\limits_{{x} \in {\Phi _i}\backslash {x_{0}}} {L_i(||x||)H^{{x}}_i{P_i}}.
\end{align}

Therefore, the signal-to-interference-plus-noise-ratio~(SINR) of $i$-th tier $\Upsilon_i$ at the typical user is given by
\begin{align}
\Upsilon_i= \frac{{L_i(||x_0||)G_0{N_i}|{h_i}{|^2}{P_i}}}{{\sigma_i^2 + I_i}},
\end{align}
where $h_2$ follows independent Nakagami fading due to utilizing mmWave and the parameter $N^p_2$ of Nakagami fading is considered to be a positive integer for simplifying the analysis~\cite{6932503}. Therefore, $|h_2|^2$ is a normalized Gamma random variable. On the other side, we assume a Rayleigh fading model for the macro tier so that the fading parameter $N^p_1\buildrel \Delta \over = 1$.
\vspace{-0.2cm}
\subsection{Cache-enabled Content Access Protocol}
\vspace{-0.1cm}
In this paper, we assume that a static multimedia content catalog containing $N_c$ files is stored at the server. Every macro BSs has a restricted storage with $M_1$ files, and each of pico BSs has a weaker ability to cache $M_2$ files, namely, $M_2< M_1< N_c$. High-speed backhauls are employed for connecting the core server to macro BSs like traditional HetNets, and the backhaul capacity is denoted by $C_{bh}$. When the data traffic load of our proposed networks becomes low, the content will be cached in a sequence of its popularity rank at all BSs via broadcasting until the storage is fully occupied. The content with $f$-th popularity rank can be represented by the Zipf distribution as shown below
\begin{align}\label{Zipf}
{P_{f}} = \frac{{{f^{ - \delta }}}}{{\sum\nolimits_{n = 1}^{{N_c}} {{n^{ - \delta }}} }},
\end{align}
where $\delta\geq0$ is the skew parameter of the probability distribution.

With the aid of Max-RP scheme, the access protocol in our paper is defined as follows.

\emph{Access Protocol:} When the typical user demands a multimedia file, it communicates with the proximate macro or pico BS with maximum received power. However, if the demanded content is absent from the corresponding BS due to limited storage capacity, the typical user will request that file from the server via the nearest macro BS.
\vspace{-0.2cm}
\section{Laplace Transform of Interference}
\vspace{-0.1cm}
The expected value of interference can be derived by \emph{Laplace Transform of Interference}, using which the success probability and ASE for the considered HetNets will be deduced.
\vspace{-0.2cm}
\subsection{Laplace Transform of Interference Analysis in Tier 1}
\vspace{-0.1cm}
In the macro tier, we utilize sub-6 GHz as the carrier frequency, and the fading channel is assumed to be Rayleigh fading. Therefore, the exact expression for Laplace transform of interference can be expressed as below.
\begin{lemma}\label{lemma1}
The Laplace transform of interference in the macro tier is given by
\begin{align}
{\mathcal{L}_1}(s) = \exp \bigg( { - \pi {\lambda _1}{r^2}\Big( {{}_2{F_1}\big( { - \frac{2}{{{\alpha _1}}},1;1 - \frac{2}{{{\alpha _1}}}; - \frac{{s }}{{{r^{{\alpha _1}}}}}} \big) - 1} \Big)} \bigg),
\end{align}
where $_2{F_1}(.)$ is the Gauss hypergeometric function.
\begin{IEEEproof}
\emph{The proof procedure is similar as \emph{Appendix B} in~\cite{liu2017caching}, but fading parameter equals to 1 in our case due to Rayleigh fading assumption.}
\end{IEEEproof}
\end{lemma}
\vspace{-0.2cm}
\subsection{Laplace Transform of Interference Analysis in Tier 2}
\vspace{-0.1cm}
 Since the path loss exponent of LOS links $\alpha_2$ is no less than 2 for the practical mmWave networks, we will divide the analysis into two conditions ($\alpha_2>2$ and $\alpha_2=2$) in order to achieve several closed-form equations.
\begin{lemma}\label{lemma2}
Under the condition $\alpha_2>2$, the $n$-th Laplace transform of interference in the pico tier is as follows
\begin{align}
\mathcal{L}_{2}^n( {s } ) =& \exp \Big( { - \pi {\lambda _2}( {R_L^2 - {r^2}} )} \nonumber\\
&{- \frac{{{\pi ^2}{\lambda _2}}}{{2u_1}}\sum\limits_{k_1 = 1}^{u_1} {{W_n}\big( {\frac{{{x_{k_1}}d}}{\lambda },s } \big)} \sqrt {( {1 - x_{k_1}^2} )} } \Big),
\end{align}
where
\begin{align}\label{a1}
{W_n}\left( {\omega ,s} \right)=&{}_{\rm{2}}{F_1}\left( { - \frac{2}{{{\alpha _{\rm{2}}}}},N_{\rm{2}}^p;1 - \frac{2}{{{\alpha _{\rm{2}}}}}; - \frac{{ns{G_2}(\omega )}}{{N_2^p{r^{{\alpha _2}}}}}} \right){r^2} \nonumber \\
 &-{}_{\rm{2}}{F_1}\left( { - \frac{2}{{{\alpha _{\rm{2}}}}},N_{\rm{2}}^p;1 - \frac{2}{{{\alpha _{\rm{2}}}}}; - \frac{{ns{G_2}(\omega )}}{{N_2^pR_L^{{\alpha _2}}}}} \right)R_L^2,
\end{align}
$x_{k_1}=\cos(\frac{{2{k_1}- 1}}{{2{u_1}}}\pi ),k_1=1,2,...,u_1,$ are Gauss-Chebyshev nodes over $[-1,1]$, and $u_1$ is a tradeoff parameter between the accuracy and complexity~\cite{7812773,7445146}. When $u_1\rightarrow \infty$, the equality is established.

Numerous actual channel measures~\cite{deng201528,rappaport201238} have indicated that the path loss exponent of LOS link is $2$ under various carrier frequencies, e.g. $28$ GHz, $38$ GHz and $73$ GHz, so we are more interested in the Laplace transform of interference under the condition of $\alpha_2=2$. The equation~\eqref{a1} will be changed into
\begin{align}\label{a1_2}
&{W_n}( {\omega ,s } ) =\frac{{ns{G_2}( \omega  ) }}{{{N^p_2}}}\Big( {{F_y}\big( {\frac{{ns{G_2}( \omega  ) }}{{{N^p_2}R_L^2}}} \big) - {F_y}\big( {\frac{{ns{G_2}( \omega  ) }}{{{N^p_2}{r^2}}}} \big)} \Big),
\end{align}
where
\begin{align}
{F_y}( y ) = &N_2^p\ln \big( {1 + \frac{1}{y}} \big) - \frac{1}{{y{{( {1 + y} )}^{N_2^p - 1}}}} \nonumber\\
&- \sum\limits_{m = 1}^{N_2^p - 1} {\frac{{N_2^p}}{{{{( {1 + y} )}^{N_2^p - m}}( {N_2^p - m} )}}} .
\end{align}
\begin{IEEEproof}
\emph{See Appendix A.}
\end{IEEEproof}
\end{lemma}
\vspace{-0.5cm}
\section{Success Probability and Area Spectral Efficiency Analysis}\label{Success_ASE}
\vspace{-0.1cm}
Before discussing the success probability, we define our \emph{Content Placement} as follows: All BSs in different tiers will choose the most popular contents to store. As a result, the probability $ p_{i,f} $ of the condition that $i$-th tier BSs cache the same copy of $f$-th ranked file can be expressed as below
\begin{align}
{p_{i,f}} = \mathbf{U}(f - 1) - \mathbf{U}(f - {M_i} - 1).
\end{align}

Therefore, a set of $i$-th tier BSs containing the $f$-th ranked file $\Phi_{i,f}$ will form an independent non-homogeneous PPP with the density ${p_{i,f}}{\lambda _i}$

From the costumer's perspective, the success probability is an important parameter to appraise the quality of service. In our cache-enabled HetNets, the instantaneous data rate at the typical user exceeding the pre-decided rate threshold $R_{th}$ will contribute to the success probability~\cite{liu2017caching}. As discussed in the previous sections, we conclude that the proposed system has two different processes in sending the multimedia contents: 1)~User Association Mode; and 2) Server Mode. We will discuss these two modes in details as below.
\vspace{-0.2cm}
\subsection{User Association Mode}
\vspace{-0.1cm}
The typical user will choose the macro or pico BS with maximum received power as its serving BS. We define the Max-RP association probability when the typical user connects with $i$-th tier as follows
\begin{align}
{\mathcal{A}^P_i} = \mathbb{P}\left[ {{N_i}{C_i}{P_i}{r^{ - {\alpha _i}}} > {N_j}{C_j}{P_j}{r^{ - {\alpha _j}}}} \right],
\end{align}
where $j\neq i$ and $j\in [1,2]$

Therefore, the probability density function~(PDF) of the distance $r$ between the typical user and its serving BS with $f$-th ranked file in $i$-th tier is changed into~\cite{jo2012heterogeneous}
\begin{align}\label{distance_distribution}
f_{i,f}^P( r ) = 2\pi {p_{i,f}}{\lambda _i}r\exp ( { - \pi \sum\limits_{j = 1}^2 {{p_{j,f}}{\lambda _j}{{\hat P}_j}^{\frac{2}{{{\alpha _j}}}}{r^{\frac{2}{{{{\hat \alpha }_j}}}}}} } ),
\end{align}
where ${{\hat P}_j} = \frac{{N_jC_j{P_j}}}{{N_iC_i{P_i}}}$ and ${{\hat \alpha }_j} = \frac{{{\alpha _j}}}{{{\alpha _i}}}$.
\vspace{-0.2cm}
\subsection{Server Mode}
\vspace{-0.1cm}
In server mode, the nearest Macro BS will act as the relay to retransmit the multimedia file from the server to the typical user. Under this condition, the
backhaul capacity will restrict the performance of our system. To simplify the notation, we first derive the coverage probability of tier 1 without considering the cache capacity.
\begin{lemma}\label{lemma3}
Since in various articles~\cite{7493676,liu2017caching}, the noise can be ignored in the traditional cellular networks with sub-6~GHz, we only consider the signal-to-interference-ratio~(SIR) instead of SINR for analyzing the performance of tier 1, namely, $\sigma _1^2=0$. Under this condition, the coverage probability for no-caching tier 1 is given by
\begin{align}
{P_{{\Upsilon _1}}}(\tau ){ = _2}{F_1}{\left( { - \frac{2}{{{\alpha _1}}},1;1 - \frac{2}{{{\alpha _1}}}; - \tau } \right)^{{\rm{ - 1}}}}.
\end{align}
\begin{IEEEproof}
\emph{Note that the SIR coverage probability ${{P}_{{\Upsilon _{1}}}}( \tau  ) = \mathbb{P}\left[ {|{h_1}{|^2} > \frac{{\tau  {{I_1} }{r^{{\alpha _1}}}}}{{{N_1}{C_1}{P_1}}}|r = ||{x_0}||} \right]$ where $|h_1|^2 \sim \exp (1)$ due to Rayleigh fading assumption. Thus such coverage probability can be expressed as ${P_{{\Upsilon _1}}}(\tau ) = \int_0^\infty  {{\mathcal{L}_1}({r^{{\alpha _1}}})} p_{1}(r)dr$. Note that $\int_0^\infty  {r\exp ( - a{r^2})dr = } \frac{1}{{2a}},a > 0$, the coverage probability will be simplified as above.}
\end{IEEEproof}
\end{lemma}
\begin{corollary}\label{corollary1}
The success probability ${P_S}( {{R_{th}}} ) $ in server mode is given by
\begin{align}
{P_S}({R_{th}}) = \mathbf{U}({C_{bh}} - {R_{th}})\sum _{f = {M_1} + 1}^{{N_c}}{P_f}{P_{{\Upsilon _{1}}}}({2^{\frac{{{R_{th}}}}{{{B_1}}}}} - 1).
\end{align}
\begin{IEEEproof}
\emph{In server mode, if $C_{bh}$ is smaller than $R_{th}$, the ${P_S}( {{R_{th}}} ) $ will be zero as the system rate is not large enough for transmitting the content. Moreover, if $C_{bh}$ is larger than $R_{th}$, the success probability for $f$-th ranked file under this mode will be ${P_f}{P_{{\Upsilon _{1}}}}( {{2^{\frac{{{R_{th}}}}{{{B_1}}}}} - 1} )$. We sum up the $(M_1+1)$-th to $N_c$-th elements to calculate ${P_S}( {{R_{th}}} ) $ as shown above.}
\end{IEEEproof}
\end{corollary}
\vspace{-0.2cm}
\subsection{Success Probability}
\vspace{-0.1cm}
Based on the Laplace transform of the interference, we are able to calculate the content-related coverage probabilities of two tiers, which is the basement for analyzing the success probability.
\begin{lemma}\label{lemma4}
For $f$-th ranked content, the coverage probability for $i$-th tier ${\Theta _{i,f}}( \tau  )$ under Max-RP scheme is shown below
\begin{align}
{\Theta _{1,f}}(\tau ) = &\int_0^\infty  {{\mathcal{L}_1}({r^{{\alpha _1}}})} f_{1,f}^P(r)dr,\\
{\Theta _{2,f}}( \tau  ) \approx &\frac{{\pi {R_L}}}{{2u_2}}\sum\limits_{n = 1}^{{N^p_2}} {{{( { - 1} )}^{n + 1}}{ N^p_2 \choose n }}\nonumber\\
&\times {\sum\limits_{k_2 = 1}^{u_2} {{F_R}\Big( {\frac{{( {{x_{k_2}} + 1} ){R_L}}}{2} } \Big)} \sqrt {( {1 - x_{k_2}^2} )} } ,
\end{align}
where
\begin{align}
{F_R}(r) = {\mathcal{L}}_2^n(\frac{{{\eta _L}{r^{{\alpha _2}}}}}{{{G_0}}}\tau )\exp ( - \frac{{n{\eta _L}{r^{{\alpha _2}}}\tau \sigma _2^2}}{{{P_2}{C_2}{N_2}{G_0}}})f_{2,f}^P(r).
\end{align}
\begin{IEEEproof}
\emph{Note that the probability of distance under this user association is shown in equation~\eqref{distance_distribution}, with the similar calculation process with \emph{Appendix A} from~\cite{6932503} and \emph{Lemma~\ref{lemma3}}, coverage probability for two tier can be expressed as above.}
\end{IEEEproof}
\end{lemma}

As mentioned in the beginning of this section, the universal success probability can be defined as below
\begin{align}
{{\cal P}_s}({R_{th}}) = \sum\limits_{i = 1}^2 {{\cal A}_i^P \mathbb{P}\left[ {{B_i}{{\log }_2}(1 + {\Upsilon _i}) > {R_{th}}} \right]}  + {P_S}({R_{th}}).
\end{align}

\begin{theorem}\label{theorem1}
With the aid of \emph{Corollary~\ref{corollary1}} and \emph{Lemma~\ref{lemma4}}, the success probability of Max-RP will be as follows
\begin{align}
{\mathcal{P}_P}( {{R_{th}}} ) \approx \sum\limits_{f = 1}^{{M_1}} {\sum\limits_{i = 1}^2 {{P_f}} {\Theta _{i,f}}( {{2^{\frac{{{R_{th}}}}{{{B_i}}}}} - 1} )}  + {P_S}( {{R_{th}}} ).
\end{align}
\begin{IEEEproof}
\emph{When the required rate is $R_{th}$, the pre-decided SINR threshold is $({{2^{\frac{{{R_{th}}}}{{{B_i}}}}} - 1} )$, so the success probability for $f$-th ranked content in the first $M_1$ popularity rank is $ {P_f} {\Theta _{i,f}}( {{2^{\frac{{{R_{th}}}}{{{B_i}}}}} - 1} ) $. Considering the whole multimedia contents, the success probability will be calculated as above.}
\end{IEEEproof}
\end{theorem}
\vspace{-0.2cm}
\subsection{Area Spectral Efficiency}
\vspace{-0.1cm}
The ASE is the average instantaneous data rate transmitted in unit bandwidth and unit area. Assuming that Gaussian Codebooks are utilized for all transmissions, we are able to define ASE with the aid of Shannon's Capacity Formula. It is expressed as follows $ASE = \lambda {\log _2}\left( {1 + \tau } \right){P_{\tau}}$, where $P_{\tau}$ is the SINR coverage probability of the considered networks and $\lambda$ denotes the active BSs' density~\cite{7446343}.

\begin{proposition}
The ASE $\mathbb{A}$ for Max-RP user association strategy is given by
\begin{align}\label{ASE}
\mathbb{A}= \sum\limits_{f = 1}^{{M_1}} {\sum\limits_{i = 1}^2 {\frac{{{P_f}{p_{i,f}}{\lambda _i}{R_{th}}}}{{{B_i}}}} } \Theta_{i,f} ({2^{\frac{{{R_{th}}}}{{{B_i}}}}} - 1) + \frac{{{\lambda _1}{R_{th}}}}{{{B_1}}}{P_S}({R_{th}}),
\end{align}
\end{proposition}
\vspace{-0.7cm}
\section{Numerical Results}
\vspace{-0.1cm}
The general network settings for our system are shown in Table.~\ref{table1}~\cite{7493676,liu2017caching,6932503} and the reference distance for the intercept is one meter.
\begin{table}[h]
\centering
\caption{General Settings of the Network}
\label{table1}
\begin{tabular}{|l|l|}
\hline
   LOS ball range     & $R_L=200$ m\\ \hline
   Density of tier1     & $\lambda_1=1/ (250^2\pi)$ m$^{-2}$\\ \hline
   Density of tier2      &$\lambda_2=20/ (250^2\pi) $ m$^{-2}$\\ \hline
   Bandwidth    & $B_1=20$ MHz; $B_2=1$ GHz\\ \hline
   Path loss law     & $\alpha_1=4$;  $\alpha_2=2$, $N^p_2=3$\\ \hline
   Number of antennas     & $N_1=1$; $N_2=16$\\ \hline
   Carrier frequency for two tiers     & $f_{m}=2$ GHz; $f_{p}=28$ GHz\\ \hline
   Transmit Power at BSs         & $P_1=80$ dBm; $P_2=30$ dBm\\ \hline
   Transmit Power at the typical user       & $P_0=30$ dBm\\ \hline
   Backhaul capacity      & $C_{bh}=500$ Mbps\\ \hline
   Number of content & $M_1=80$, $M_2=10$, $N_c=100$\\ \hline
   Skew of the popularity distribution & $\delta=0.6$\\ \hline
\end{tabular}
\end{table}

\begin{figure}
  \centering
  \includegraphics[width= 3.2 in, height=1.85 in]{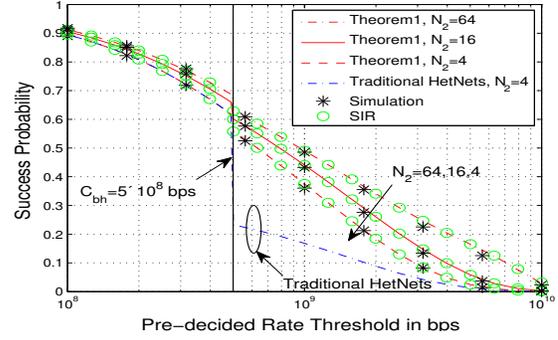}\\
  \caption{Success probability versus pre-decided rate threshold with $B_1=500$~MHz.}\label{fig1}
\end{figure}

Fig.~\ref{fig1} illustrates the impact of $R_{th}$ and $N_2$ on success probability. Comparing the analytical results of the success probability with the simulation results, we note that they match each other ideally, thereby certifying the analysis. It is obvious that our cache-enabled HetNets outperform the traditional HetNets where the macro BSs have no caching capacity, especially in the area $R_{th}\geq C_{bh}$. Moreover, the SIR scenario fits the analytical results perfectly so that our model can be regarded as an interference-limited system. Lastly, with the rise of antenna scales $N_2$, the success probability will correspondingly increase.
\begin{figure}
  \centering
  \includegraphics[width= 3.2 in, height=1.85 in]{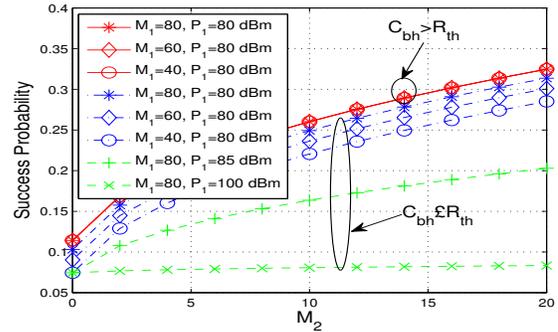}\\
  \caption{Success probability versus cache capacity $M_2$ in tier 2 with different $M_1$, $P_1$ and $R_{th}=10^8$ bps.}\label{fig2}
\end{figure}
\begin{figure}
  \centering
  \includegraphics[width= 3.2 in, height=1.85 in]{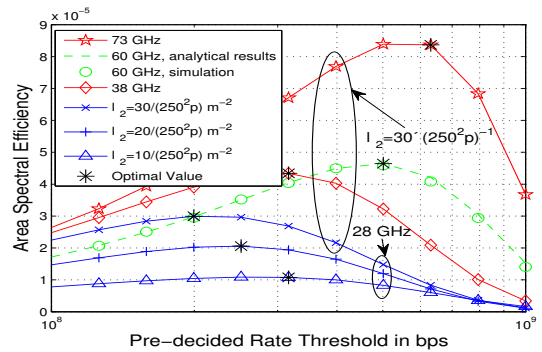}\\
  \caption{Area spectral efficiency versus pre-decided rate threshold with different $\lambda_2$, carrier frequencies and $B_1=500$ MHz.}\label{fig3}
\end{figure}

Fig.~\ref{fig2} shows the success probability versus cache capacity with different $P_1$. The success probability has a negative correlation with the transmit power $P_1$. Massive $P_1$ contributes to large received power, when $P_1$ increases to 100 dBm, the success probability becomes independent on $M_1$. In this case, the typical use will associate with tier 1 all the time. In terms of the cache capacity, the success probability is a monotonic increasing function with pico BSs' storage capacity $M_2$. For the cache capacity of macro BSs $M_1$, when $C_{bh}>R_{th}$, the success probability has no relationship with $M_1$ since the less-popular content that only contained in the server can be transmitted freely through the backhaul. Under this condition, the proposed HetNet is same with the traditional one. On the other side, when $C_{bh}\leq R_{th}$, the server will be blocked so that the success probability will be benefited by the large $M_1$, which represents that more multimedia files can be stored at macro BSs.

 Fig~\ref{fig3} plots the ASE versus $R_{th}$ with different $\lambda_2$ and carrier frequencies. The optimum pre-decided rate threshold $R_{th}$ for achieving the maximum ASE can be easily figured out due to the convex property. When the density of pico tier $\lambda_2$ increases from $10/(250^2\pi)$ m$^{-2}$ to $30/(250^2\pi)$ m$^{-2}$, the optimal number decreases. Then, we compare the carrier frequencies at 28 GHz, 38 GHz, 60 GHz and 73GHz with antenna scales $N_2=$10, 20, 40 and 80~\cite{8016632}, respectively. All of them has the same LOS path loss exponents ($\alpha_2=2$), except that the path loss exponent for 60 GHz is 2.25~\cite{deng201528,rappaport201238}. Fig.~\ref{fig3} illustrates that 73 GHz is the best choice thanks to the largest antenna scales. The simulation of 60~GHz matches the analytical results with an insignificant difference, thereby verifying all analytical expressions under the condition $\alpha_2>2$.
\vspace{-0.3cm}
\section{Conclusion}
\vspace{-0.2cm}
In this treatise, we utilize the stochastic geometry to analyze the performance of our cache-enabled hybrid HetNet. More specifically, the proposed network, which performs better than the traditional HetNet, can be regarded as an interference-limited system due to the high density of mmWave tier and the nature of sub-6~GHz tier. As discussed in the previous sections, our system has a positive correlation with pico BSs' antenna scales and the cache capacity of both tiers. Additionally, there exists an optimum value of pre-decided rate threshold contributing to the maximum ASE. Lastly, for Max-RP association strategy, 73 GHz is the best carrier frequency of mmWave tier.
\vspace{-0.5cm}
\numberwithin{equation}{section}
\section*{Appendix~A: Proof of Lemma~\ref{lemma1}} \label{appendixA}
\renewcommand{\theequation}{A.\arabic{equation}}
\setcounter{equation}{0}
\vspace{-0.2cm}
The Laplace transform of interference in tier 2 is given by
\begin{align}
&\mathcal{L}_2^n( {s } ) = \mathbb{E}\Big[ {\exp \big( { - ns \sum\limits_{x \in {\Phi _2}\backslash x_0} {{G_2}( \omega  )|{h_2}{|^2}||x|{|^{ - {\alpha _2}}}} } \big)} \Big] \nonumber\\
\mathop  = \limits^{(a)} & e^{  { - 2\pi {\lambda _2}{\mathbb{E}_{{G_2}}}\Big[ {\int_r^{{R_L}} {\big( {1 - {{\big( {1 + \frac{{ns {G_2}( \omega  )}}{{N_2^p{v^{{\alpha _2}}}}}} \big)}^{ - N_2^p}}} \big)vdv} } \Big]} } \nonumber\\
\mathop  = \limits^{(b)} &e^{  { - \frac{{\pi {\lambda _2}\lambda }}{d}}{\int_{ - \frac{d}{\lambda }}^{\frac{d}{\lambda }} {\int_r^{{R_L}} {\big( {1 - {{\big( {1 + \frac{{ns {G_2}( \omega  )}}{{N_2^p{v^{{\alpha _2}}}}}} \big)}^{ - N_2^p}}} \big)vdv} d\omega } } },
\end{align}
where (a) follows the Gamma random variable's moment generating function~\cite{6932503}; (b) is computing the expectation of tier 2 antenna gain $G_2$. When $\alpha_2>2$, (A.1) can be simplified as follows
\begin{align}
&\mathcal{L}_2^n( {s } ) \mathop  = \limits^{(c)}\exp \bigg( - \pi {\lambda _2}(R_L^2 - {r^2}) \nonumber \\
&- \frac{{\pi {\lambda _2}\lambda }}{{2d}}\int_{ - \frac{d}{\lambda }}^{\frac{d}{\lambda }} {\Big({}_{\rm{2}}{F_1}\left( { - \frac{2}{{{\alpha _{\rm{2}}}}},N_{\rm{2}}^p;1 - \frac{2}{{{\alpha _{\rm{2}}}}}; - \frac{{ns{G_2}(\omega )}}{{N_2^p{r^{{\alpha _2}}}}}} \right){r^2}} \nonumber\\
& - {}_{\rm{2}}{F_1}\left( { - \frac{2}{{{\alpha _{\rm{2}}}}},N_{\rm{2}}^p;1 - \frac{2}{{{\alpha _{\rm{2}}}}}; - \frac{{ns{G_2}(\omega )}}{{N_2^pR_L^{{\alpha _2}}}}} \right)R_L^2\Big)d\omega \bigg),
\end{align}
(c) follows  Gauss hypergeometric function. When $\alpha_2=2$, (A.1) can be simplified as below
\begin{align}
\mathcal{L}_2^n( {s } ) \mathop  = \limits^{(d)} &\exp{\Big({  { - \pi {\lambda _2}( {R_L^2 - {r^2}} )}}}- \frac{{\pi {\lambda _2}\lambda }}{{2d}}\int_{ - \frac{d}{\lambda }}^{\frac{d}{\lambda }} {\big( {\frac{{ns{G_2}( \omega  ) }}{{N_2^p}}}}\nonumber\\
 &\times {{{{{\big( {{F_y}( {\frac{{ns{G_2}( \omega  ) }}{{N_2^pR_L^2}}} ) - {F_y}( {\frac{{ns{G_2}( \omega  ) }}{{N_2^p{r^2}}}})}\big)} \big)d\omega } } }\Big)}.
\end{align}
(d) follows (2.117-1), (2.117-3) and (2.118-1) in~\cite{jeffrey2007table}. With the aid of  Gauss-Chebyshev Quadrature, we obtain \textbf{Lemma~\ref{lemma1}}.
\vspace{-0.2cm}
\bibliographystyle{IEEEtran}
\bibliography{mybib}

\end{document}